\documentclass[aps,prd,twocolumn,showpacs,superscriptaddress,footinbib]{revtex4-1}  % for review and submission
\usepackage{graphicx}  % needed for figures
\usepackage{dcolumn}   % needed for some tables
\usepackage{bm}        % for math
\usepackage{comment}
\usepackage{amssymb}   % for math
\usepackage{amsmath}
\usepackage[usenames,dvipsnames]{color}
\usepackage[normalem]{ulem}

\newcommand{\pau}[1]{}

\definecolor{chmagenta}{rgb}{0.0, 0.5, 1.0}

\begin{document}

\preprint{APS/123-QED}

  \title{Post-Newtonian Dynamics in Dense Star Clusters:
 Formation, Masses, and Merger Rates of Highly-Eccentric Black Hole Binaries}

  \author{Carl L.\ Rodriguez}
  \affiliation{Pappalardo Fellow; MIT-Kavli Institute for Astrophysics and Space Research, 77 Massachusetts Avenue, 37-664H, Cambridge, MA 02139, USA}

       \author{Pau Amaro-Seoane}
          \affiliation{
Institute of Space Sciences (ICE, CSIC) \& Institut d'Estudis Espacials de Catalunya (IEEC)\\
at Campus UAB, Carrer de Can Magrans s/n 08193 Barcelona, Spain\\
Institute of Applied Mathematics, Academy of Mathematics and Systems Science, CAS, Beijing 100190, China\\
Kavli Institute for Astronomy and Astrophysics, Beijing 100871, China            }

  \author{Sourav Chatterjee}
         \affiliation{Tata Institute of Fundamental Research, Department of Astronomy and Astrophysics, Homi Bhaba Road, Navy Nagar, Colaba, Mumbai, 400005, India}

  \author{Kyle Kremer}
         \affiliation{Center for Interdisciplinary Exploration and Research in
    Astrophysics (CIERA) and Dept.~of Physics and Astronomy, Northwestern
      University, 2145 Sheridan Rd, Evanston, IL 60208, USA}

        \author{Frederic A.\ Rasio}
         \affiliation{Center for Interdisciplinary Exploration and Research in
    Astrophysics (CIERA) and Dept.~of Physics and Astronomy, Northwestern
      University, 2145 Sheridan Rd, Evanston, IL 60208, USA}

    \author{Johan Samsing}
         \affiliation{Department of Astrophysical Sciences, Princeton University, Peyton Hall, 4 Ivy Lane, Princeton, NJ 08544, USA}

        \author{Claire S.\ Ye}
         \affiliation{Center for Interdisciplinary Exploration and Research in
    Astrophysics (CIERA) and Dept.~of Physics and Astronomy, Northwestern
      University, 2145 Sheridan Rd, Evanston, IL 60208, USA}
      
              \author{Michael Zevin}
         \affiliation{Center for Interdisciplinary Exploration and Research in
    Astrophysics (CIERA) and Dept.~of Physics and Astronomy, Northwestern
      University, 2145 Sheridan Rd, Evanston, IL 60208, USA}

\date{\today}% It is always \today, today,

\begin{abstract}

 Using state-of-the-art dynamical simulations of globular clusters, including radiation reaction during black hole encounters and a cosmological model of star cluster formation, we create a realistic population of dynamically-formed binary black hole mergers across cosmic space and time.   We show that in the local universe, 10\% of these binaries form as the result of gravitational-wave emission between unbound black holes during chaotic resonant encounters, with roughly half of those events having eccentricities detectable by current ground-based gravitational-wave detectors.  The mergers that occur inside clusters typically have lower masses than binaries that were ejected from the cluster many Gyrs ago.  Gravitational-wave captures from globular clusters contribute 1-2 $ \rm{Gpc}^{-3}\rm{yr}^{-1}$ to the binary merger rate in the local universe, increasing to $\gtrsim 10$  $\rm{Gpc}^{-3}\rm{yr}^{-1}$ at $z\sim 3$.  Finally, we discuss some of the technical difficulties associated with post-Newtonian scattering encounters, and how care must be taken when measuring the binary parameters during a dynamical capture.  

%\begin{description}
%\end{description}
\end{abstract}

\maketitle
% * <fred.rasio@gmail.com> 2017-12-11T14:53:46.866Z:
% 
% "scattering" to many people means flyby, not strong encounter
% 
% ^.
%\tableofcontents

\section{Introduction}
\label{sec:intro}

In early 2016, Advanced LIGO reported the first detection of gravitational waves (GWs) from a binary black hole (BBH) merger \cite{Abbott2016a}, demonstrating for the first time the feasibility of GW astronomy.  With the subsequent detection of four more BBH mergers \cite{Abbott2017e,Abbott2017c,Abbott2017,Abbott2016} and one binary neutron star inspiral \cite{Abbott2017d}, we are rapidly approaching an era where catalogues of GWs will supplant single detections.   While individual events can provide significant physical insight (particularly when coupled to electromagnetic observations), the power of GW astrophysics will soon come from comparing entire populations of compact-object mergers to detailed astrophysical models.

A significant amount of work has been done to try and understand the various possible formation pathways for merging BBHs.  Broadly, most formation channels fall into one of two categories: isolated binary stellar evolution, where the BBH is formed as the remnant of a massive stellar binary \cite{Belczynski2002,Voss2003,
Podsiadlowski2003,Sadowski2007a,Belczynski2010,Dominik2012,Dominik2014,Dominik2013,Belczynski2016}, and dynamical formation, where the BBH is created and hardened through dynamical interactions in a dense stellar environment \cite{PortegiesZwart2000,Banerjee2010,Ziosi2014,Banerjee2017,
OLeary2006,OLeary2007,Moody2009,Downing2010,Downing2011,Tanikawa2013,Bae2014,
Rodriguez2015a,Rodriguez2016a,Rodriguez2016b,Askar2016,Giesler2017,Rodriguez2018,Banerjee2017,Hong2018,Choksi2018,Fragione2018}.  Both of these broad categories can produce BBHs with masses, spins, and merger rates consistent with the current LIGO/Virgo constraints \cite[e.g.,][]{Amaro-Seoane2016}, and there is no reason to suspect that multiple channels do not operate simultaneously, making the question of their origins particularly challenging.

However, one key observable that is unique to the dynamical formation channels is the potentially high orbital eccentricity of BBHs as they enter the sensitivity band of the GW detectors.  Historically, eccentricity has not been included in many of the studies of compact-object mergers, because the emission of GWs can efficiently circularize a binary long before it reaches the frequency band of ground-based detectors.  While this is certainly true for BBHs formed from isolated stellar binaries, many dynamical channels can produce BBHs mergers that retain significant eccentricities even up to a GW frequency of $10\,$Hz or greater.  This can either arise from the influence of a third bound object, which effectively ``pumps up'' the binary eccentricity via the Lidov-Kozai mechanism \cite{Antonini2012a,Antonini2016,
VanLandingham2016,Leigh2017,Silsbee2017,Petrovich2017,Hoang2018,Antonini2017,Rodriguez2018a,Arca-Sedda2018}, or by directly forming the BBHs with very high eccentricities near the detection threshold for terrestrial detectors \cite{Samsing2014,OLeary2009,Samsing2017a,Samsing2017,Amaro-Seoane2016,Samsing2018,Samsing2017b,Rodriguez2018,Zevin2018}.  While the latter is nearly impossible in the purely Newtonian case, the inclusion of post-Newtonian (pN) corrections to the equations of motion offers a new pathway for BBH formation.  In particular, the 2.5pN term describing non-conservative radiation reaction allows for the formation of bound BBHs by the emission of GWs {\it during\/} a close BH encounter \cite[e.g.,][]{Hansen1972,Quinlan1989}.   These encounters are most likely to occur during three- and four-body scatterings, where the long-lived chaotic states offer many opportunities for unbound BHs to pass sufficiently close that they emit a pulse of GWs and form a highly-eccentric binary \cite[e.g.,][]{Gultekin2006,Samsing2014}.   These encounters occur frequently in the cores of dense stellar environments such as globular clusters (GCs), where a binary will continue to encounter other stars and binaries until it is either disrupted, ejected from the cluster, or merges due to GW emission \cite[e.g.,][]{Benacquista2013}, 

Recently, semi-analytic \cite{Samsing2018} and fully-numerical \cite{Rodriguez2018} models of GCs have shown that these GW-driven captures can contribute significantly to the BBH merger rate, with roughly 5\% of sources from dense star clusters entering the LIGO/Virgo detection band with an eccentricity of 0.1 or higher.  These sources emit GWs with a unique spectral signature, providing a telltale sign of dynamical formation.  In parallel, much work has been done to develop waveforms that can model these unique signatures, placing the measurement of orbital eccentricity from BBH mergers within the grasp of Advanced LIGO/Virgo \cite{Mikoczi2015,Huerta2018,Hinder2010,Tanay2016,Gondan2017,Moore2016,Huerta2017,Yunes2009,Pierro2001,Sperhake2008,Huerta2014}.  The latest generation of waveforms, combining pN theory with numerical relativity and BH perturbation theory, are approaching sufficient fidelity that they can enable both the detection and characterization  of eccentric mergers \cite{Huerta2018}.  To avail ourselves of this new, dynamically-rich parameter space, we must better understand the formation, evolution, and mergers of eccentric BBHs across cosmic space and time.

In this paper, we expand upon our previous work \cite{Rodriguez2018}, to fully explore the formation, masses, and merger rates, of eccentric BBH mergers from GCs.  Using state-of-the-art dynamical models of GCs, along with a recently developed cosmological model for cluster formation \cite{El-Badry2018,Rodriguez2018b}, we perform a complete population survey of eccentric BBH mergers from the cores of dense star clusters.  In Section 2, we describe the changes to our method from \cite{Rodriguez2018}, and how we combine the cosmological model for GC formation from \cite{El-Badry2018} with our $N$-body models of GC evolution to produce a realistic population of BBH mergers across cosmic time.  In Section 3, we explore the various mechanisms by which BBH mergers are produced in GCs, and describe the expected eccentricities and masses for each sub-channel, both globally and in the local universe ($z<1$), as well as the formation of BBH mergers with GW frequencies \emph{inside} the LIGO/Virgo band.  Finally, in Section 4, we explore the cosmological rate of eccentric BBH mergers from GCs.  Throughout this paper, we assume a
flat $\Lambda \rm{CDM}$ cosmology with $h= 0.679$ and $\Omega_M = 0.3065$
\cite{PlanckCollaboration2015}, and that all BHs from stars are born with no spin (though they can attain spin through mergers with other BHs), which allows for the retention of BH merger products \cite[see e.g.,][]{Rodriguez2018} in the cluster. 

\section{Methods}

\subsection{Monte Carlo Models}
\label{sec:mccosmo}

We generate our GC models using the Cluster Monte Carlo (\texttt{CMC)} code, a H\'enon-stlye \cite{Henon1971,Henon1971a} Monte Carlo code for modeling the evolution of massive, spherical star clusters.  \texttt{CMC} has been developed over many years \cite{Joshi1999,Pattabiraman2013}, and can model all of the relevant physical processes which drive the evolution of dense star clusters.  This includes orbit-averaged two-body relaxation, single and binary
stellar evolution \cite{Hurley2000,Hurley2002,Chatterjee2010}, three-body binary formation from single BHs
\cite{Morscher2012,Morscher2015}, three- and four-body scattering encounters between stars and black holes \cite{Fregeau2004,Fregeau2007}, physical collisions, and galactic tides.  This approach has been shown to produce GC models similar to those generated with direct $N$-body calculations \cite[e.g.][]{Wang2016} in a fraction of the computational time \cite{Rodriguez2016}.  \texttt{CMC} uses prescriptions for stellar winds, supernovae, pulsational-pair instabilities, and BH formation \cite{Rodriguez2016a,Rodriguez2018} which are identical to the single stellar evolution prescriptions used in the most recent population synthesis approaches to modeling BBHs formed from massive stellar binaries \cite{Belczynski2010,Fryer2012,Dominik2013,Belczynski2016,Belczynski2016a}.  \texttt{CMC} also includes pN corrections to encounters between BHs and BBHs using the technique developed in \cite{Antognini2014,Amaro-Seoane2016,Rodriguez2018}, allowing us to self-consistently model GW emission and BBH mergers during strong resonant encounters.  

We consider a grid of 48 GC models, covering a wide range of initial 
conditions. As in 
\cite{Rodriguez2018}, we created 24 GC models spread across a 4x3x2 grid in initial mass, metallicity, and virial radius. 
Our models span four initial particle numbers ($2\times10^6$, $10^6$, $5\times10^5$, and $2\times10^5$) 
corresponding roughly to initial masses of $1.2\times10^6M_{\odot}$, 
$6\times10^5M_{\odot}$, $3\times10^5M_{\odot}$, and $10^5M_{\odot}$, three separate stellar metallicities ($0.01Z_{\odot}$, 
$0.05Z_{\odot}$, and $0.25Z_{\odot}$) at different galactocentric radii (20 kpc, 8 kpc, and 2 kpc respectively, roughly following the correlation in the Milky Way), and two initial virial radii (1 pc and 2 pc).  The masses of single stars and the primary masses of binaries are chosen from a Kroupa initial mass function \cite{Kroupa2003} in a range between $0.08 M_{\odot}$ to $150M_{\odot}$.  For each of these models, we assume 10\% of our particles are initially in binaries, with semi-major axes drawn from a distribution flat in the log from stellar contact to the local hard/soft boundary, eccentricities drawn from a thermal distribution ($p(e)de= 2ede$), and secondary masses drawn from a uniform distribution between 0 and the primary mass.  We assume that the binary fraction is independent of the stellar mass.  In addition to the 4x3x2 grid of models, we generate 4 additional, statistically-independent realizations for each of the models with $2\times10^5$ initial particles.  This was done to increase the statistics for mergers from smaller clusters (some of which may disrupt before the present cosmological era), where any individual cluster may only produce a few (if any) BBH mergers over its lifetime.  This yields a total number of 48 GC models.  

In previous works employing realistic GC models \cite[e.g.,][]{Downing2011,Rodriguez2016a,Askar2016}, it was assumed that all GCs formed at the same epoch, i.e. that all GCs formed exactly 12 Gyr ago at redshift 3.5.  However, this well-known approximation ignores the measured spread in GC ages and the correlation between age and cluster metallicity \cite[e.g., ][]{Forbes2010}.  Furthermore, it ignores the continued formation of dense star clusters up to the present day (e.g.\ super-star clusters), and the correlation between formation times and galaxy halo mass \cite[though this was done correctly in][]{Chatterjee2017}.  Here, we extend our GC models to redshift 10, corresponding to an age of 13.3 Gyr, and convolve their time evolution with a cosmological model for GC formation which we now describe.

Comparing collisional models of GC evolution to 
detailed cosmological simulations is well-beyond both the scope of this paper and the resolution of most cosmological simulations.  However, recent models have been developed that can predict the formation and evolution of GCs across cosmic times and galaxy type, either using halo merger trees and detailed prescriptions for gas-cloud collapse \cite[e.g.][]{Boylan-Kolchin2017,Choksi2018,El-Badry2018}, or semi-analytic treatments of cluster evolution and disruption in the galaxy \cite[e.g.][]{Fragione2018a}.  These models have been used to estimate the rates of various transients from dense star clusters \cite[e.g.][]{Hong2018,Fragione2017,Fragione2018,Rodriguez2018b,Choksi2018a} while properly accounting for the formation, destruction, and evolution of said clusters across cosmic time.

\begin{figure}[tb]
\centering
\includegraphics[scale=0.85, trim=0in 0.in 0in 0.in, clip=true ]{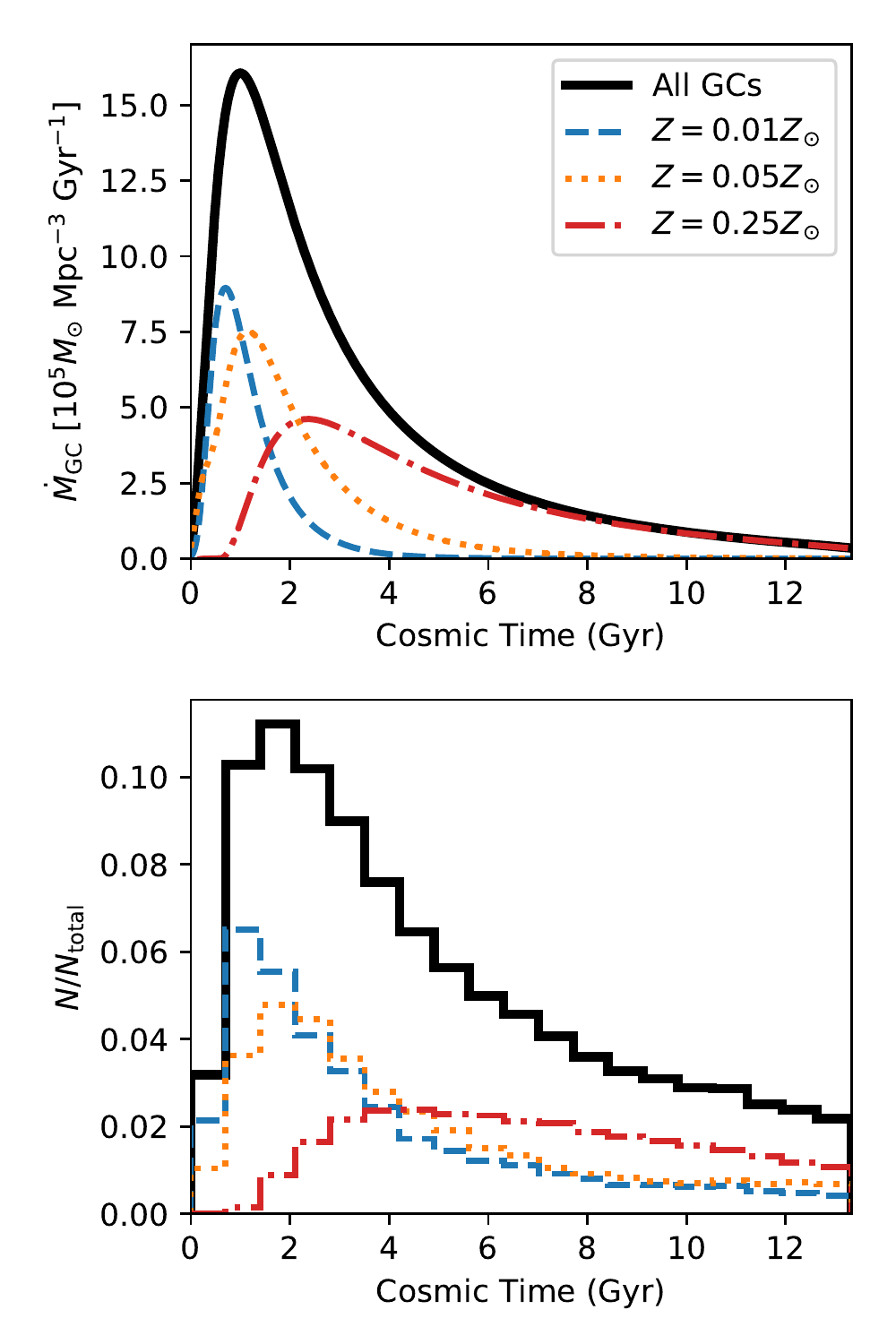}
\caption{Formation rate of GCs as a function of cosmic time (where $t=0$ at $z=10$).  In the top panel, we show the rate of GC formation over all galaxy halo masses, based on the model of \cite{El-Badry2018}, and the three metallicity bins that we use to assign individual ages to our GC models.  On the bottom panel, we show the overall merger rate of BBHs from our GC models, convolved with the GC formation rate.  The numbers are normalized to the total number of BBH mergers from clusters of all metallicities.}
\label{fig:formation}
\end{figure}

In \cite{Rodriguez2018b}, we used the GC formation model from \cite{El-Badry2018} to estimate the BBH merger rate from GCs formed in different halo masses across cosmic times.  {This was done using the GC models presented in here.}  We extend this formalism, using it both to draw our initial distribution of cluster ages and to determine how to weigh the BBH mergers from the 48 cluster models to best represent a realistic population of GCs.  To that end, we use the fitting formula from \cite{Rodriguez2018b}, to Figure 5 of \cite{El-Badry2018} which captures the formation rate of GCs per galaxy halo mass at a given redshift (accounting for the halo mass function).  See Appendix A of \cite{Rodriguez2018b}.  Integrating this rate directly over all halo masses (from $10^8M_{\odot}$ to $10^{14}M_{\odot}$) would give us the distribution of formation times for GCs.  However, we can improve upon this simple estimate by binning the integral according to the median metallicity of star formation occurring in halos of a certain mass at a given redshift.  To do this, we place our GC models in the center of three logarithmically spaced bins in metallicity space; i.e., we assume our $0.01Z_{\odot}$ clusters represent all clusters formed with $Z < 0.022 Z_{\odot}$, the $0.25Z_{\odot}$ clusters cover the bin where $Z > 0.11 Z_{\odot}$, and the $0.05Z_{\odot}$ clusters represent clusters in the middle bin.  To determine the formation rate for a specific metallicity, we only integrate over halo masses where the median gas metallicity at a given redshift for that halo mass lies within the metallicity bin of interest:

\begin{equation}
\dot{M}_{\rm GC}(\tau) = \int_{\left<M_{\rm Halo}\right>(Z_{\rm low})}^{\left<M_{\rm Halo}\right>(Z_{\rm high})} \left. \frac{ \dot{M}_{\rm GC}} {d\log_{10}M_{\rm    
	 Halo}} \right|_{z(\tau)} dM_{\rm Halo}
	 \label{eqn:elbad}
	\end{equation}

\noindent The median star formation metallicity for a given halo mass at a given redshift is taken from \cite{Behroozi2013}, and the relation between stellar metallicity and gas metallicity is taken from \cite{Ma2016}.  This is identical to the stellar enrichment prescription used in \cite{El-Badry2018}, and we find it to produce similar results.

We show the distribution of formation times for clusters with specific metallicities in Figure \ref{fig:formation}.  As would be expected from any reasonable cosmological model, the higher metallicity clusters form at later time, and all cluster formation in the local universe occurs in high-metallicity clusters.  In the bottom panel of Figure \ref{fig:formation}, we show the distribution of merger rates from our cluster models, convolved with the cluster formation times calculated above.  Even though they represent a smaller total fraction of clusters in the local universe, the continuous formation of clusters leads to the high-metallicity GCs playing the largest contribution to the BBH mergers in the local universe.  To generate the population of BBH mergers that we will use for the remainder of this paper, we assign to each BBH merger 100 unique formation times from Figure \ref{fig:formation}, allowing us to convolve the BBH merger rate with the cosmological model from \cite{El-Badry2018}.  

\subsection{Weighting the GC Models}

To get a representative sample of BBH mergers from GCs, we 
must apply a weighting scheme to our grid of models, in order to better represent the observed and theoretically-predicted properties of GCs.  In \cite{Rodriguez2016a}, this was accomplished by binning the 
population of GCs in mass/metallicity space, then drawing with replacement from our 
sample of BBH mergers according to the weight of each model (i.e.\ models with 
higher weights contributed more BBHs to our effective population in the local 
universe).  Here, we can take a more cosmologically-motivated approach.  We assign to each 
model a weight based on the cluster initial mass function (CIMF) and 
metallicity.  We assume that GCs form following a $1/M^2$ distribution, then 
assign to each cluster a weight based on its initial mass, assuming that 
the cluster occupies the center of a linearly-spaced bin in the CIMF.  The least 
massive clusters ($\sim10^5M_{\odot}$) get a weight of $\sim 50\%$, while 
the most massive clusters ($\sim1.2\times10^6M_{\odot}$) contribute only 
$\sim 10\%$ of their binaries.  Of course, many of our smallest clusters 
disrupt before the present day, meaning that the contribution from these 
clusters to the merger rate in the local universe will either arise from clusters that were formed at late times or from BBHs which were ejected from their parent clusters prior to disruption.  Additionally, to each model we assign a metallicity weight.  This is 
simply the fraction of GCs that formed in each metallicity bin from 
Figure \ref{fig:formation}.  The weight assigned to each BBH merger is simply the product of 
the mass and metallicity weights assigned to each cluster. 

Throughout the rest of this paper, any quantities we quote (averages, 
histograms, cumulative fractions, etc.) will be weighted 
according to this scheme, except where otherwise noted.  We also note that there 
still remains uncertainty in the CIMF, especially given observational evidence of an
exponential-like truncation at higher masses in young massive star clusters \cite[e.g.,][]{PortegiesZwart2010}.  However, to test the sensitivity 
of our results on the CIMF, we also considered a weighting scheme based on the 
present-day GC mass function (GCMF), taken by shifting the peak the observed 
luminosity function of GCs \citep{Harris2014} upwards by a factor of 4 
to account for a mass-to-light ratio of 2 \cite{Bell2003} and the (roughly factor of 2) mass loss 
experienced by GCs over their $\sim12$Gyr lifetimes \cite[e.g.][]{Morscher2015}.  
Because the most massive clusters contribute the majority of sources, and 
because the number of BBH mergers within a Hubble time from a single cluster 
scales super-linearly with the cluster mass \cite[e.g.][]{Rodriguez2016a}, the 
most massive clusters from the current grid still dominate the BBH mergers in the local universe in either case; however, we 
find only minimal changes in the properties of BBH mergers when using the CIMF 
versus the observed GCMF.  Therefore, we only report the results using the CIMF for the remainder of this work.

\subsection{pN Integrations of BH Encounters}
\label{sec:pnfix}

We have made several changes to the pN implementation described in \cite{Rodriguez2018}, which we will now detail.  First, for the runs described in this work, we only integrate BH encounters using purely Newtonian forces plus the 2.5pN term, which is responsible for the emission of GWs.  This represents a departure from \cite{Amaro-Seoane2016,Rodriguez2018}, where both the 1pN and 2pN terms were included as well{, but is consistent with the models used for the cosmological rate estimate in \cite{Rodriguez2018b}}.  While including all terms up to and including the 2.5pN terms obviously includes more physics, this can introduce significant errors in the classification of bound systems in the \texttt{Fewbody} integrator.

By default \texttt{Fewbody} reclassifies the entire hierarchical structure of the system every 500 timesteps, to determine when to terminate any encounter and how to report the outcome.However, this classification scheme depends on the energy and angular momentum 
of bound systems in the encounter, which are converted into orbital 
elements (semi-major axis and eccentricity). With the inclusion of pN terms, the Newtonian energy and angular momentum are no 
longer formal constants of the motion (even without the dissipative 2.5pN 
terms), and can vary significantly over a single orbital period.  Furthermore, 
because the pN expansion is only valid up to the next highest order in 
$(v/c)$, there do not exist well-defined constants of the motion that can be used to identify particles and bound or unbound.  Even the pN definitions of the energy and angular momentum are only conserved in the orbit-averaged approximation.  For a binary whose component velocities are not constant over the orbit (such as the highly-eccentric systems studied here), the pN energy and angular momentum can vary significantly over a single orbit.   

\begin{figure}[htb]
\centering
\includegraphics[scale=0.85, trim=0.0in 0in 0in 0in, clip=true ]{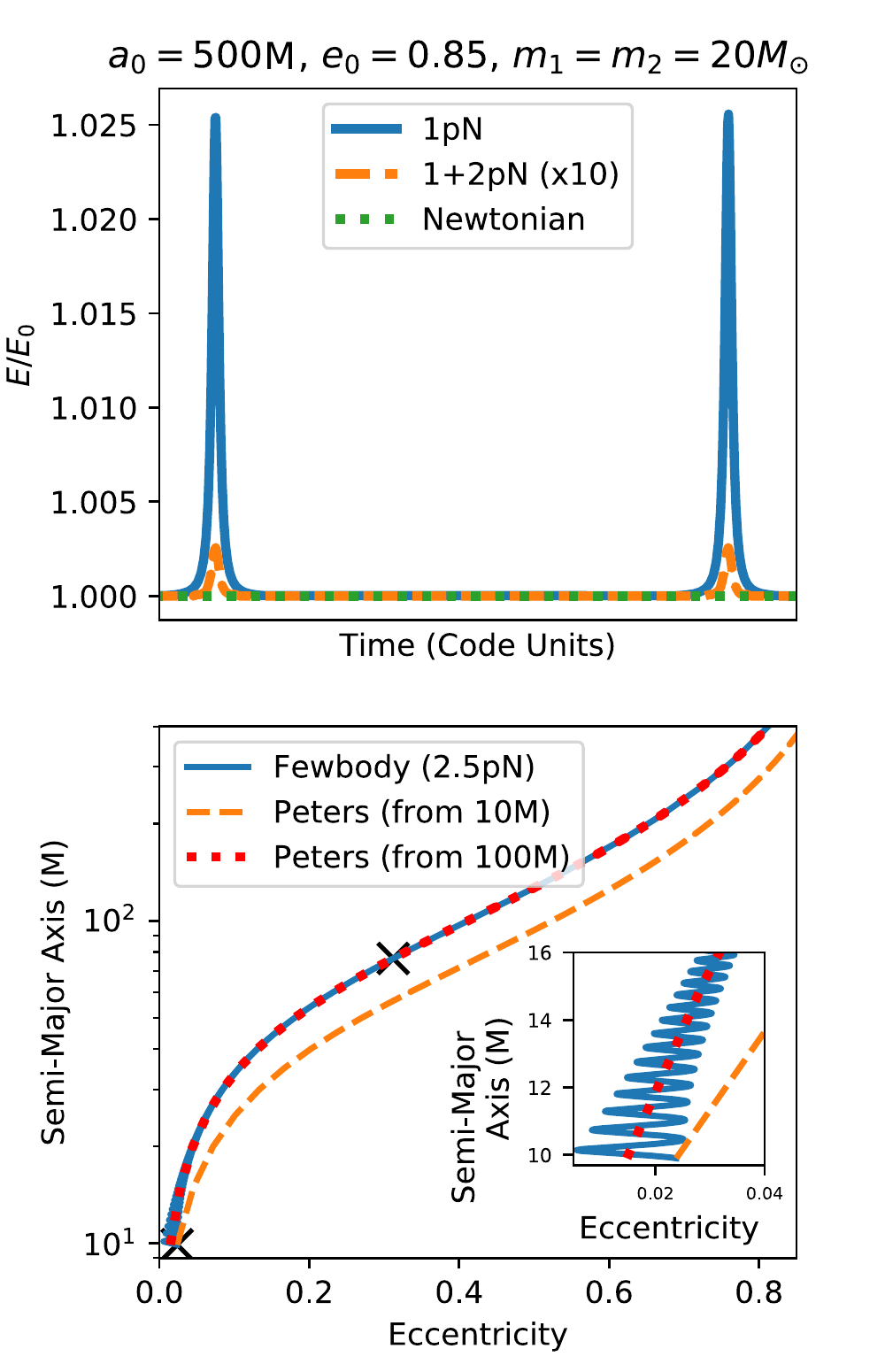}
\caption{A pN integration of an isolated binary with close separation and high eccentricity, typical of GW capture BBH mergers with high eccentricities.  In the top panel, we show the energy of the binary over two orbits when only the conservative terms are included.  Because the pN expansion explicitly does not conserve energy beyond the highest order in $v/c$, the binary introduces a significant energy increase during the high-velocity pericenter passage.  This error does not occur in the purely Newtonian case.  In the bottom panel, we show the variation in the Newtonian eccentricity near merger.  The integration of Equations \eqref{eqn:dadt} and \eqref{eqn:dedt} from initial conditions extracted at instantaneous separations of $10M$ and $100M$ are shown in dashed orange and dotted red, respectively.}
\label{fig:Fewbody}
\end{figure}

In the top panel of Figure \ref{fig:Fewbody}, we show the fractional change in 
the energy over a single orbit of a moderately eccentric binary ($m_1 = m_2 = 
20M_{\odot}$, $e_0 = 0.85$ and $a_0 = 500M$, where $M$ is the total mass of the binary in geometric units, i.e.\ $(m_1 + m_2) G/c^2$).  As the velocity changes 
from apocenter to pericenter by a factor of $\sim 12$, the  energy increases by nearly 2.5\% when only 1pN corrections are considered.  When both 1 and 2pN terms are included, the change in energy decreases by nearly two orders of magnitude.  Note that we are using the pN definitions of the energy \citep[e.g.][]{Mora2004} corresponding to each pN order.  We find that the fractional change in energy from apocenter to pericenter is directly proportional to the next-to-next-to leading order pN correction.  In other words, the change in the 1pN energy is proportional to $(v/c)^6$, while the 2pN energy difference scales as $(v/c)^8$. 
While the increase in the example in Figure \ref{fig:Fewbody} is relatively minor (especially when 2pN terms are included), this change can be extreme for the GW captures we are interested in, which frequently occur at the $e \rightarrow 1$ boundary between unbound and bound systems.  In these cases, the variation in the energies can be much more extreme, and for binaries where $(v/c) \sim 0.1$ at pericenter, the variation over a single orbit can be as much as 1000\% (10\%) for the 1pN (2pN) equations of motion.  Because of these issues, we choose to remove the 1 and 2pN terms from our integration in this work.  This decreases the number of mergers with eccentricities of $\sim 10^{-3}$ at a GW frequency of 10Hz that were reported in \cite{Rodriguez2018}, which arose from the classification of unbound systems as highly-eccentric (and bound) binaries.  

{When considering the long-term dynamics of triple systems, particularly hierarchical triples that undergo Kozai-Lidov oscillations, the contributions from the conservative pN terms can be substantial, since the relativistic precession of the binary pericenter can suppress (or in rare cases enhance) the eccentricity growth of the inner binary \cite{Naoz2013}.  This can be particularly important when considering the formation of long-lived triples in dynamical environments, which may contribute at the $\sim 1\%$ level in GCs \cite{Antonini2016}, or even higher in open clusters \cite{Banerjee2018a}.  CMC does not currently track the secular evolution of hierarchical systems, and so we do not consider these effects here.} But for the scattering encounters that dominate the mergers presented here, the conservative 1 and 2pN terms do not appear to have a significant impact.  Recent work \cite{Zevin2018} has shown that the inclusion of conservative pN terms during scattering encounters such as these do not have any noticeable influence on the statistical outcomes of these scattering experiments.  This, combined with the difficulties in correctly measuring the energies of these systems, was the primary motivation for our exclusion of the conservative pN terms. {However, because the inclusion of higher-multiplicity BH systems would only increase the number of highly-eccentric BBH mergers, we consider our results here to be a conservative lower limit.}

{Finally, we note that an error was discovered in the pN physics from our previous work \cite{Rodriguez2018} that changed the strength of the relativistic terms for scattering encounters with non-equal mass components.  This error did not significantly alter our previous results (the fraction of mergers occurring in the cluster decreased by $\sim 10\%$ in the local universe) but we mention it here for consistency.  This error was discovered in \cite{Rodriguez2018b}, and does not effect the rate estimates quoted there.}

\subsection{Measuring Eccentricity during GW Captures}

To better measure the eccentricities of any BBHs that merge during resonant encounters, we also sample the orbital elements of merging BHs at multiple discrete separations.  We record the semi-major axes and eccentricities for all pairs of BHs (bound or unbound) when they approach within 500M, 100M, 50M, and 10M, where M is the total mass of the BH pair.  To determine when a binary crosses into the LIGO/Virgo band, we must know the eccentricity and semi-major axis when that binary's GW frequency passes 10 Hz.  For a circular binary, the GWs are all emitted at the lowest-order harmonic ($n=2$) of the orbital frequency, such that the GW frequency is simply twice the orbital frequency.  However, an eccentric binary emits GWs across a range of harmonics of the orbital frequency.  The dominant frequency at which an eccentric binary emits GWs can be approximated as \cite[][]{Wen2003}

\begin{equation}
	f_{\rm GW} = \frac{\sqrt{G M}}{\pi} \frac{(1 + e)^{1.1954}}{\left[ 
	a(1-e^2)\right]^{3/2}}~,
	\label{eqn:wen}
\end{equation}

\noindent where $a$ and $e$ are the semi-major axis and eccentricity, respectively.   To determine when a binary will enter the LIGO/Virgo band, we can simply integrate the orbit-averaged equations for the evolution $a$ and $e$ from \cite{Peters1964}:

\begin{align}
\left<\frac{da}{dt}\right> &= -\frac{64}{5} \frac{G^3 m_1 m_2 (m_1+m_2)}{c^5 a^3 (1-e^2)^{7/2}} \left( 1+\frac{73}{24}e^2 + \frac{37}{96}e^4 \right)~,\label{eqn:dadt}\\
\left<\frac{de}{dt}\right> &= -\frac{304}{15} e \frac{G^3 m_1 m_2 (m_1+m_2)}{c^5 a^4 (1-e^2)^{5/2}} \left( 1 + \frac{121}{304} e^2 \right)~,\label{eqn:dedt}
\end{align}

\noindent until Equation \eqref{eqn:wen} equals 10Hz.  Here $m_1$ and $m_2$ are the component masses of the binary.  

To ensure that we are measuring the correct $a$ and $e$ to use as initial conditions for the Peters equations, we pick the largest $(a,e)$ pair sampled from $10M$ to $500M$ where the system is bound, in order to minimize the error introduced by sampling the binary in the strong-field regime.  These Newtonian orbital elements can vary over an orbital timescale, particularly in the strong-field regime near merger (see the bottom panel of Figure \ref{fig:Fewbody}). Note that there are two cases where this procedure may not provide a solution: first, it is possible that sources can form \emph{inside} the LIGO/Virgo band, at such a close separation that their dominant frequency is already greater than 10Hz when the binary forms. This typically occurs when the eccentricity is extremely large, so we simply report these systems as having eccentricities of 0.99 at 10Hz.  Secondly, we find that a handful of BBHs are technically unbound when they merge in a \texttt{Fewbody} encounter.  However, because we treat two BHs as having merged whenever they pass within 10M of one another, we cannot resolve whether these systems were the result of true collisions between unbound particles, or whether they were binaries that formed with pericenters less than 10M.  We record these systems as having eccentricities $e\gtrsim 1$, and will discuss both cases in Section \ref{sec:pericenter}.

\section{Eccentricities of Merging Binaries}

For the purposes of this study, we will divide our BBH mergers into four categories \citep[see also][]{Zevin2018}:

\begin{itemize}
    \item \textbf{Primordial Binaries} - mergers that occur though isolated binary stellar evolution in the GC.  Of course, BBHs from binary stars can undergo dynamical encounters in the cluster before merging.  Here we label a binary as primordial only if it never participated in a strong dynamical encounter.  
    \item \textbf{Ejected Mergers} - BBHs undergoing many hardening encounters before being ejected from the cluster, only to merge later in the field \cite[e.g.][]{Downing2011,Rodriguez2016a}  
    \item \textbf{In-Cluster Mergers} - binaries merging inside the GC after a dynamical encounter, but {\it not\/} due to significant GW emission {\it during\/} the encounter.   
    \item \textbf{GW Captures} - BHs merging {\it during\/} a resonant encounter due to strong GW emission at a very close passage.

\end{itemize}

For the remainder of this paper, we will focus on the ejected, in-cluster, and GW capture channels, since these are unique to the dynamical environments in which they are formed.  

\begin{figure}[tb!]
\centering
\includegraphics[scale=0.85, trim=0in 0.175in 0in 0in, clip=true ]{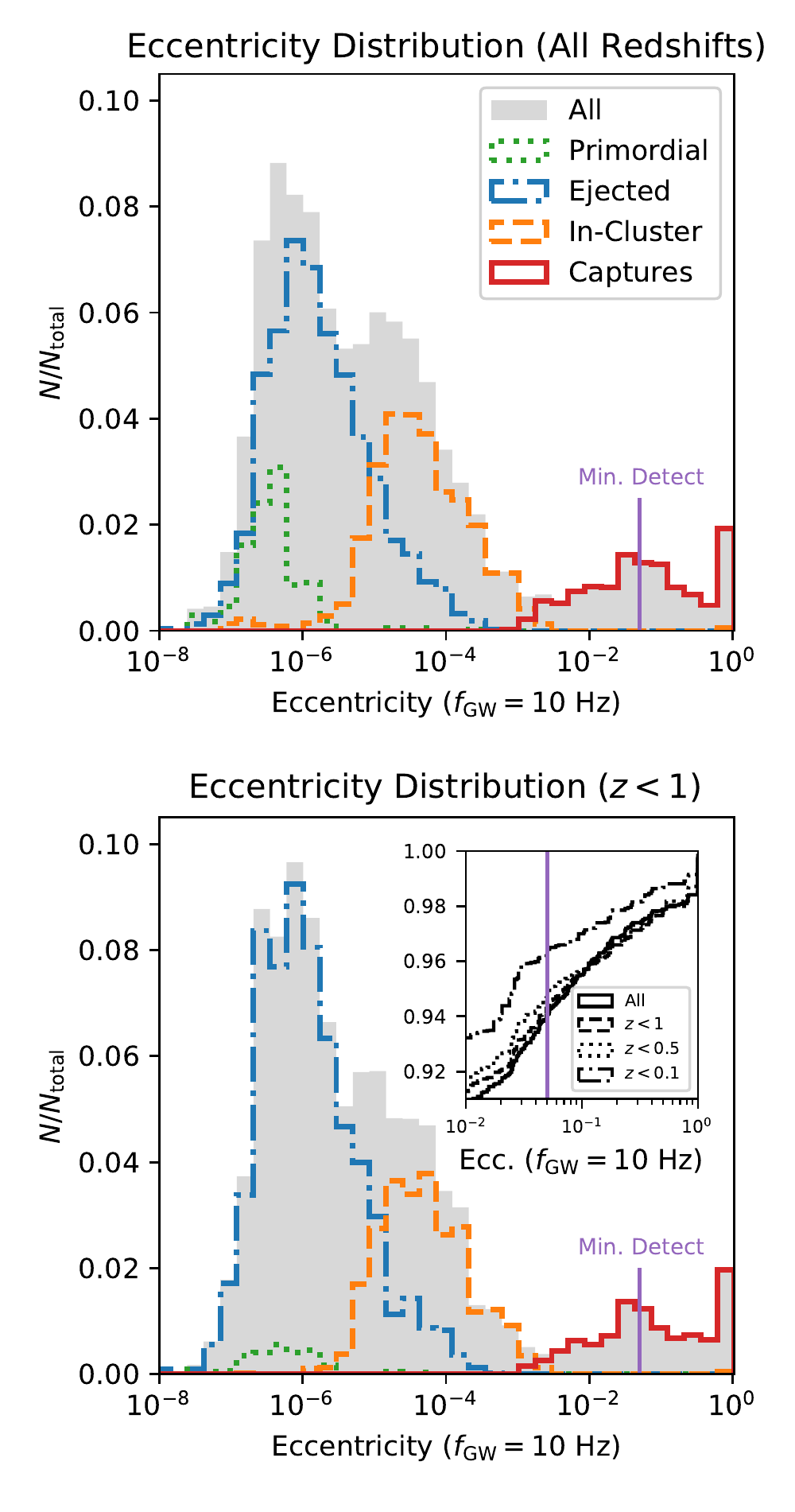}
\caption{The eccentricities at a GW frequency of 10Hz for all BBHs from GCs.  We show separately the eccentricities for the primordial binaries (BBHs from pre-existing stellar binaries, in dotted green), the BBHs that merge after being ejected from the cluster (in dot-dashed blue), the BBHs that merge in the cluster as isolated binaries (in dashed orange), and the binaries which merge due to GW emission during resonant three- and four-body encounters between BHs (in solid red).  Each curve is normalized to the total number of BBH mergers (in solid gray) and weighted using the cosmological model described in Section \ref{sec:mccosmo}.  The top panel shows all mergers, while the bottom panel is restricted to mergers in the local universe ($z<1$).  The insert in the bottom panel shows the cumulative distribution of eccentricities for all BBH mergers from GCs at different redshifts.  In each plot, we show the minimum measurable eccentricity of BBH mergers in Advanced LIGO/Virgo \cite[$e\sim 0.05$, from ][]{Lower2018}}
\label{fig:ecc}
\end{figure}

In Figure \ref{fig:ecc}, we show the total distribution of eccentricities at a GW frequency of 10Hz from our GC models at all redshifts and in the local universe ($z<1$).  The final eccentricities span a wide range, from $e = 10^{-8}$ all the way to $e\sim 1$.  However, each of the four types of BBH mergers inhabit a distinct range of the eccentricity space.  Both the primordial binaries (which mostly circularize during mass transfer) and the ejected binaries peak near $10^{-6}$.  This is consistent with previous results that have shown that both ejected cluster binaries and BBHs from the field should have relatively similar final eccentricities \cite{Breivik2016,Amaro-Seoane2016,Nishizawa2016}, with the cluster binaries having a tail extending to higher eccentricities.  

The total fraction of mergers in each sub-population changes as a function of redshift.  For the total population (out to redshift 10) ejected BBHs are the dominant contribution, comprising 55\% of all mergers from GCs.  The remaining 45\% of mergers occur in the cluster, with 5\% resulting from primordial binaries, 28\% merging as isolated binaries in the cluster, and 12\% forming as GW captures.  In the local universe ($z < 1$), the contribution from ejected binaries increases, since many of the BBHs that were ejected early in the cluster lifetime may not merge for many Myr or Gyr (see Section \ref{sec:rates}).  At these later times, the ejected BBHs comprise 64\% of all mergers, with primordial binaries, in-cluster mergers, and GW captures contributing 1\%, 25\%, and 10\%, of the total BBH mergers, respectively.  

The red histogram in Figure \ref{fig:ecc} shows the distribution of eccentricities from GW captures.  Unlike \cite{Rodriguez2018}, the distribution ranges from as low as $10^{-3}$ at a peak GW frequency of 10Hz, all the way to $e\sim1$.  This is largely due to our more conservative physics for the pN scatterings described in Section \ref{sec:pnfix}.  While the total number of GW captures has increased significantly, the fraction of mergers with measurable eccentricities is identical to our previous results, with roughly $4\%$ of all mergers from GCs entering the LIGO/Virgo band with eccentricities greater than 0.1.  This fraction increases to $\sim6\%$ if we consider the lower threshold ($e \gtrsim 0.05$) for measurably eccentric BBHs recently proposed by \cite{Lower2018}, although we note that study focused on BBH mergers with masses similar to GW150914 ($30M_{\odot}+30M_{\odot}$), which is more massive than the typical highly-eccentric merger identified here (see Section \ref{sec:masses}).  

\subsection{Why BBHs Merge Where They Do}

Of these dynamical channels, what primarily differentiates the three?  As binaries interact with other BHs and stars in the cluster, hard binaries (those whose binding energy is greater than the typical kinetic energy of surrounding stars and BHs) will preferentially harden after each encounter, shrinking their semi-major axes and leaving the encounter with some fraction of that binding energy converted to kinetic energy.  This statistical inevitability, known as Heggie's law \cite{Heggie1975}, will continue, producing harder and harder binaries until the binary either merges (due to GW emission) or is ejected from the cluster by the recoil of the encounter.  Which of these two available pathways a binary takes is largely a matter of timescales.  After an encounter, the survival of a binary is dictated by the competition of two timescales:  the timescale for GW emission to drive a binary to merge, given by

\begin{equation}
    T_{\rm GW} \propto a^4(1-e^2)^{7/2}~,
\end{equation}

\noindent where $a$ is the semi-major axis of the binary, and $e$ is the eccentricity, and the average time between successive binary encounters, which scales as

\begin{equation}
    T_{\rm bs} \propto n a^2 \sigma \left(1+\frac{G M}{2a\sigma^2}\right)~,
\end{equation}

\noindent where $n$ is the number density of single particles and $\sigma$ is the typical velocity dispersion.  As shown by Heggie, each resonant encounter (between objects of near-equal masses) will produce binaries with eccentricities drawn from a thermal distribution, $p(e)de = 2ede$.   Because of the extreme dependence of $T_{\rm GW}$ on the orbital eccentricity, if a binary leaves any scattering encounter with a sufficiently high eccentricity, it can easily merge before being disrupted or disturbed by a third body.

\begin{figure}[tb!]
\centering
\includegraphics[scale=0.85, trim=0in 0.2in 0in 0in, clip=true ]{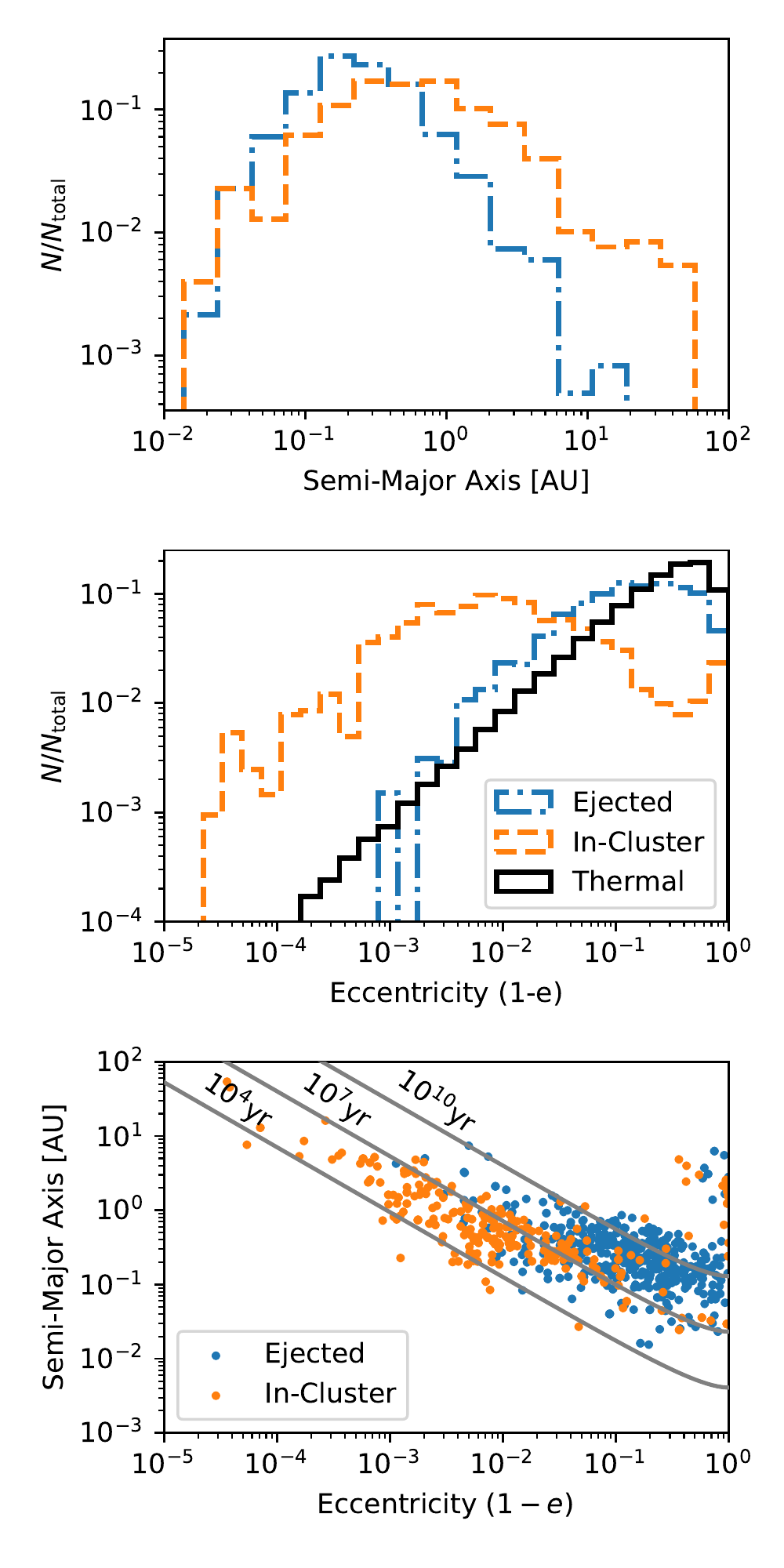}
\caption{The semi-major axes and eccentricities of the ejected and in-cluster BBH mergers immediately following their last dynamical interaction in the cluster.  The ejected mergers typically leave the cluster with smaller semi-major axes (dot-dashed blue, in the top panel) than the in-cluster mergers (dashed orange), and follow a roughly thermal eccentricity distribution (middle panel).  The in-cluster mergers typically occur when a BBH leaves a resonant encounter with a very high eccentricity, prompting a rapid merger before another BH or stellar encounter can occur.  In the bottom panel, we show jointly the semi-major axis and eccentricity for 1000 BBHs drawn from our weighted population of BBH mergers, along with the merger times for $20M_{\odot}+20M_{\odot}$ binaries at those semi-major axes and eccentricities.}
\label{fig:eccevol}
\end{figure}

In Figure \ref{fig:eccevol}, we show the post-encounter eccentricities and semi-major axes for each BBH after its last encounter in the cluster.  For ejected BBHs, this is also the encounter responsible for its ejection. Here, it is immediately obvious that the in-cluster mergers are preferentially selected from BBHs which leave their last encounters with a very high eccentricity, while the ejected BBHs leave the cluster with a distribution of eccentricities very close to thermal.  Because of the steep dependence of the inspiral time on eccentricity, the semi-major axes of the in-cluster mergers after their last encounter can be significantly larger than the ejected binaries, with a few percent of in-cluster mergers having semi-major axes in excess of 10 AU.   As an example, a $30M_{\odot}+30M_{\odot}$ binary with a separation of $a=10\rm{AU}$ and an eccentricity of $e=0.9999$ will merge in $10^4\rm{yr}$, well before another binary encounter (though this will depend on the concentration of the cluster).  This is obvious in the bottom panel of Figure \ref{fig:eccevol}.  Typically, most in-cluster mergers occur in less than $\sim 10$Myr after their last encounter in the cluster, while the ejected BBHs can may have a significant delay between their ejection and merger (up to many Hubble times).  

\subsubsection{Predicting Where BBHs Merge}

{
Each of the three dynamical channels operates on a unique timescale, $\tau$, which allows us to make significant analytic progress.  For ejected mergers, that timescale is roughly a Hubble time, $\tau_{\rm ej} \sim T_{H}$, since after ejection the BBH need only merge within the lifetime of the Universe to be of interest to LIGO/Virgo.  For in-cluster mergers, a BBH needs to merge before its next encounter in the cluster, making its maximum lifetime roughly $\tau_{\rm in}  \sim T_{\rm bs}$, while for GW captures, the timescale is roughly the orbital timescale of the binary at the beginning of an encounter ($\tau_{\rm GW} \sim T_{\rm orb}$) \citep[see e.g.,][]{2018MNRAS.481.5445S, 2018MNRAS.481.4775D}.  These timescales can be used to make specific predictions for the fraction of mergers that occur in each dynamical channel, as has been done in \cite{2018ApJ...853..140S, 2018PhRvD..97j3014S,2018MNRAS.481.5445S, 2018MNRAS.481.4775D}.  We can compare this formalism, developed by comparing the difference in timescales at each stage in a binary's life in the cluster to our numerical results, demonstrating that this simple consideration of timescales can predict the relative fraction of in-cluster mergers and GW captures from GCs.  
}

{
We start by deriving the probability, $P(t < T)$, for an assembled BBH to merge due to GWs within time $T$. Assuming the eccentricity distribution of the dynamically-assembled BBHs follows a thermal distribution $P(e)de =2ede$, one can show that \citep{2018ApJ...853..140S, 2018PhRvD..97j3014S},
}
\begin{align}
P(t < T) &\approx \left(\frac{T}{T_{\rm GW}^{e=0}}\right)^{2/7} \nonumber \\
&\propto T^{2/7} a^{-8/7}M^{6/7},
\end{align}

\noindent where $T_{\rm GW}^{e=0}$ is the GW inspiral time of a given binary (assuming a circular orbit) \cite{Peters1964}.
{
To derive the relative number of events of each outcome, we assume that BBHs form in their clusters at the hard-soft boundary, where the binding energy of the binary is equal to the average kinetic energy of nearby particles, since binaries with lower energies are typically destroyed during strong encounters \cite{Heggie1975}.  We then assume that these binaries are hardened during binary-single encounters with other BHs, with each encounter reducing the binary's semi-major axis by a fixed fraction ($\delta = 7/9$).  This continues until the binary reaches a separation, $a_{\rm ej}$, where the liberation of $\sim 20\%$ of the binding energy is sufficient to eject the BBH from the cluster.  Of course, during each encounter, there exists some probability that the binary will either merge promptly due to GW emission, or will leave with such a large eccentricity that it merges before its next BH encounter. As shown in \cite{2018PhRvD..97j3014S}, the probability
for a BBH to undergo a merger of outcome type $i$ can be written as,
}
\begin{equation}
P_{i} \approx F_{i} \times \left(\frac{\tau_{i}(a_{\rm ej})}{t_{\rm GW}^{e=0}(a_{\rm ej})}\right)^{2/7},
\end{equation}

{
\noindent where $P_{i}$ is the probability for that a BBH initially formed at the hard-soft boundary and is hardened until ejection
results in outcome $i$, and $F_{i}$ is a pre-factor based on the typical number 
of encounters a BBH undergoes in the cluster, weighted according to the GW 
merger probability of each encounter.
As shown in \cite{2018PhRvD..97j3014S}, for in-cluster mergers $F_{\rm in} \approx (7/10)/(1-\delta) \approx 3$, while for GW captures, $F_{\rm GW} \approx (7/5)/(1-\delta) \times N_{\rm MS} \approx 120$,
where $N_{\rm MS}$ is the number of intermediate, meta-stable BBH states a binary forms \textit{during} a resonant three-body encounter.
Evaluating $P_{i}$ for each outcome, assuming $a_{\rm ej} \sim 0.5$ AU, $M \sim 30M_{\odot}$, $\tau_{\rm in} \sim 10^7$ years, and $\tau_{\rm GW} \sim 0.1$ year,
we find that $P_{\rm in} \approx 0.15$ and $P_{\rm GW} \approx 0.03$. This implies that approximately 82\% of dynamically-formed BBHs
will be ejected from the cluster.
Of this ejected population a fraction $P(t_{\rm GW} (a_{\rm ej}) < T_{\rm H}) 
\approx 0.35$ will merge within $T_{\rm H}$, meaning that the probability for a 
BBH formed at the hard-soft boundary to get ejected and merge within a Hubble time is approximately $0.82 \times 0.35 \approx 0.3$. 
}{
Of course, the escape speed and central concentration of a GC can change by factors of a few over the evolution of the cluster system, changing both the rate at which BBHs are produced and their typical semi-major axes at ejection \cite[e.g.,][]{Rodriguez2016a}.  However, as a simplifying assumption we assume that BBHs are dynamically formed at the hard-soft boundary at a steady rate.  
The total contribution to the BBH merger rate from each outcome is simply the probabilities from the previous paragraph normalized to the sum of the three outcomes, i.e.~$P_{i}/\sum_{i} P_{i}$. Applying this, we find that ejected mergers, in-cluster mergers, and GW captures
contribute $62\%$, $31\%$, and $7\%$ of the total cluster merger rate, respectively. Note that for this estimate we have not included binary-binary interactions, which
have been shown to contribute about $30\%$ of all GW captures \cite{2018arXiv181000901Z}. If this were accounted for, we would expect that $\sim 10\%$ of BBH mergers would come from GW captures.  These numbers are in good agreement with the numerical results presented here, indicating that the analytic approach developed in \cite{2018PhRvD..97j3014S, 2018MNRAS.481.5445S, 2018MNRAS.481.4775D} can provide a reasonable and effective estimate of where and how BBHs merge from dense star clusters.
}
%JS:---------------------------

\subsubsection{Distinguishing In-cluster Mergers from GW Captures}

It is natural to worry that our distinction between in-cluster mergers and GW captures is entirely a result of our termination criterion for scattering encounters.  By default, \texttt{Fewbody} will end any encounter that has resolved into bound, hierarchical systems that are moving away from one another, while the pN corrections from \cite{Antognini2014} introduce an additional criteria that the pericenter velocity of any binary be less than 5\% the speed of light.  One could easily imagine, however, that binaries could be formed by the emission of GWs during an encounter and be classified by \texttt{Fewbody} as stable systems, causing the merger to be classified as an isolated, in-cluster merger.  

\begin{figure}[tb!]
\centering
\includegraphics[scale=0.85, trim=0in 0.in 0in 0in, clip=true ]{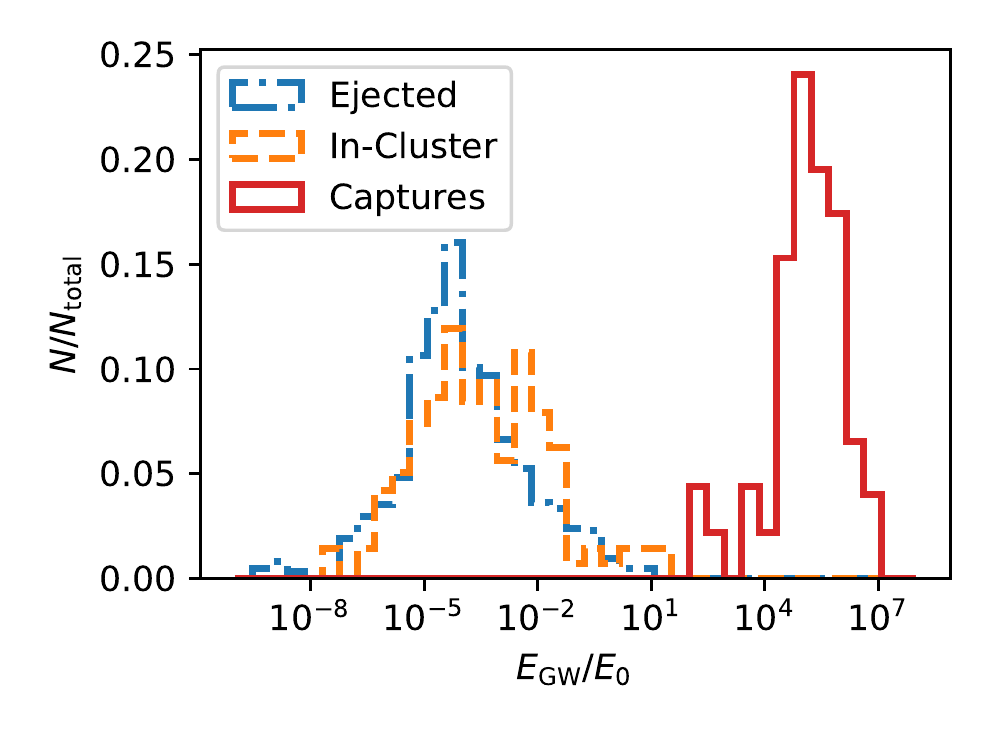}
\caption{The fraction of energy lost to GWs during a resonant three- or four-body encounter, normalized to the total initial energy (kinetic and potential) of the encounter.  The ejected and in-cluster mergers (dot-dashed blue and dashed orange) typically lose only a small fraction of their energy to GWs, while the GW capture events (in solid red) lose significant amounts of energy to GWs during the \texttt{Fewbody} encounter.}
\label{fig:egw}
\end{figure}

To test the robustness of our classification scheme, we track the GW emission during each scattering encounter by directly evaluating $\frac{dE}{dt}$ between each pair of particles in \texttt{Fewbody} \cite[e.g.,][c.f. Equation 2.10]{Iyer1995}, then summing the contributions and integrating over the total time of the encounter. In Figure \ref{fig:egw}, we show the distributions of total energies lost due to GWs during the last encounter before each merger, normalized to the total energy (kinetic + potential) at the beginning of each encounter.  Clearly, the three populations show distinct distributions in terms of energy emitted, with the GW captures showing significantly more energy loss than either the ejected or in-cluster mergers.  The ejected and in-cluster mergers, on the other hand, typically show fractional energy losses several orders-of-magnitude below unity.  We note that the in-cluster mergers show a suggestive bimodalitiy, with peaks at $E_{\rm GW}/E_0 \sim 10^{-4}$ and $\sim10^{-2}$, which does appear to weakly correlate with the eccentricity of the binary after its last encounter before merger (which we do not show).  However, our current implementation of \texttt{Fewbody} cannot discriminate between high-eccentricity systems that were created by GW emission, or binaries that were created and emitted GWs as an isolated system before the termination of the \texttt{Fewbody} integration.

\begin{figure}[tb!]
\centering
\includegraphics[scale=0.85, trim=0in 0.in 0in .in, clip=true ]{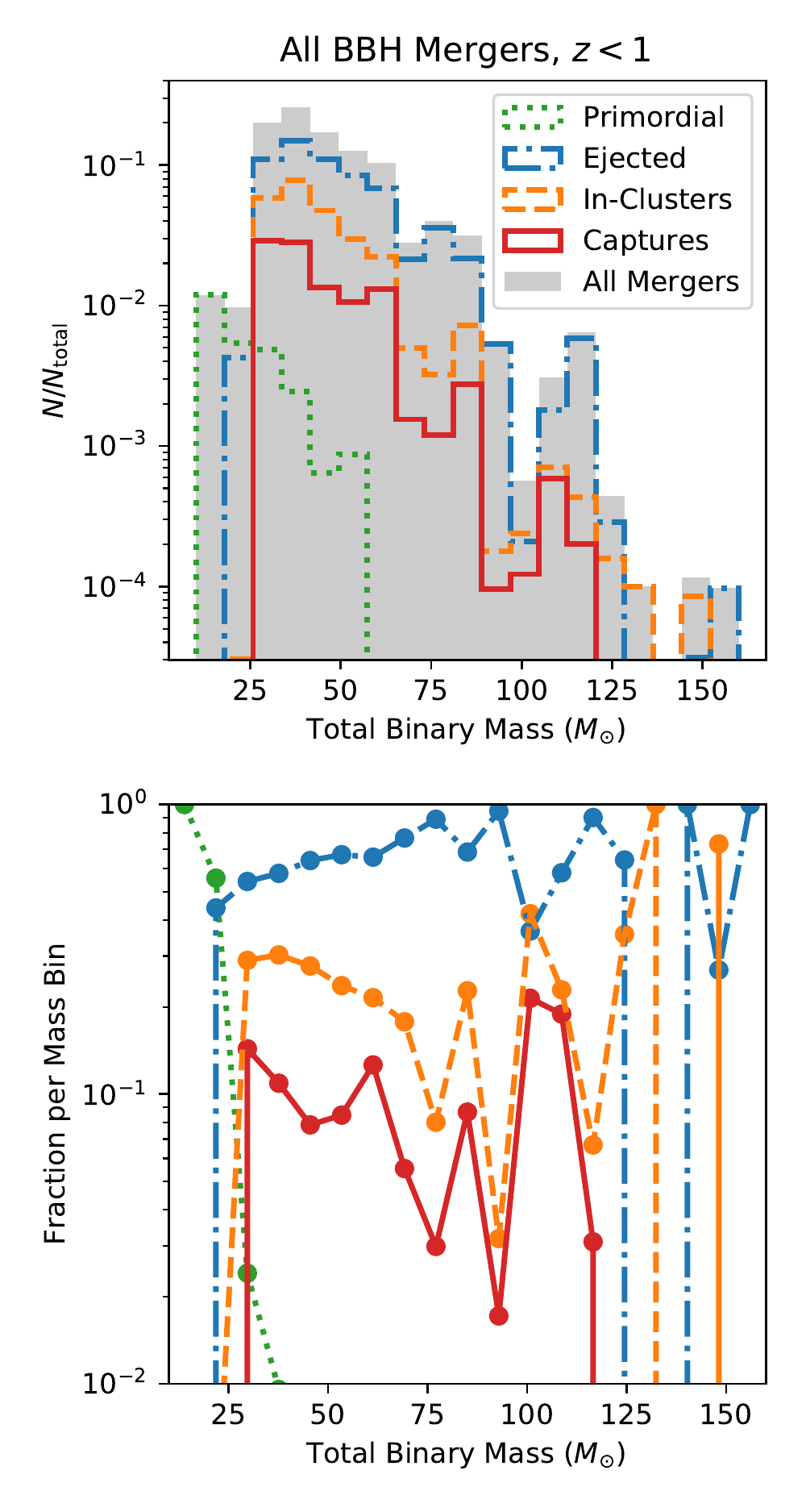}
\caption{The masses of BBH mergers from different formation pathways for all mergers that occur in the local universe ($z<1$).  In the top panel, we show the weighted total masses of the BBHs, normalized by the total number of mergers.  Although there is a trend for BBHs which merge in the cluster (both isolated binaries and GW captures) to have lower masses, this trend becomes less clear past $M \sim 80M_{\odot}$ because of the contribution  from second-generation mergers (those with at least one component formed in a previous BBH merger).  In the bottom panel, we show the fraction of mergers of each type that contribute to the total number of mergers at a given mass.}
\label{fig:masses}
\end{figure}

\begin{figure}[tb!]
\centering
\includegraphics[scale=0.85, trim=0in 0.in 0in .in, clip=true ]{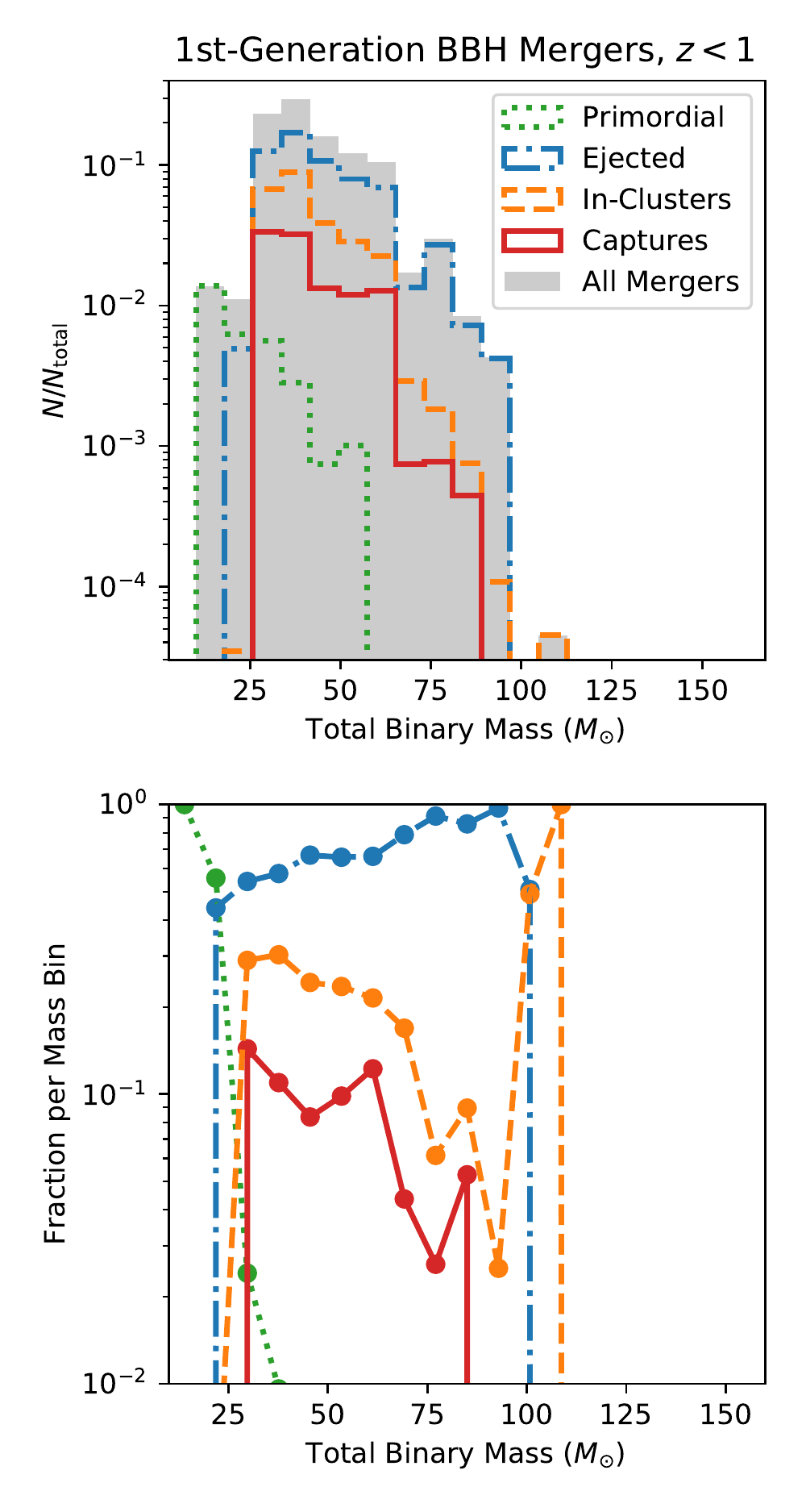}
\caption{Similar to Figure \ref{fig:masses}, but excluding any mergers whose components were formed in a previous BBH merger.  This enforces a much stronger cutoff in total mass, with only a few objects having total masses greater than $81M_{\odot}$ (the result of mass transfer onto BHs during previous star-BH encounters).  Here the trend towards ejected mergers having larger masses than in-cluster mergers and GW captures is significantly more pronounced.}
\label{fig:masses_1st}
\end{figure}

\subsection{Masses of Eccentric BBH Mergers}
\label{sec:masses}

During the evolution of a GC, the most massive BHs are preferentially ejected first, since they are able to most efficiently segregate into the center of the cluster and participate in dynamical encounters \cite[e.g.][]{Morscher2015}. By the present day, most GCs are expected to be entirely depleted of their heaviest BHs, leaving behind a population of lower mass BHs to participate in dynamical encounters.  Because of this, it would be natural to expect that the masses of BBHs that merge inside clusters in the local universe are typically lower than those that may have been ejected many Gyr ago, when the cluster still contained significantly more massive BHs.  This was in fact the result in \cite{Rodriguez2018}, which did not include any cosmological treatment of GC formation, and assumed that all GCs formed precisely 12 Gyr ago.  Here, by considering a more realistic distribution of GC formation times (Section \ref{sec:mccosmo}), we find that the difference in masses for ejected mergers, in-cluster mergers, and GW captures are not nearly as clean cut.

In the top panel of Figure \ref{fig:masses}, we show the weighted distribution of BBH masses which merge in the local ($z<1$) universe.  As with previous results \cite{Rodriguez2016a,Rodriguez2016b}, the peak of the distribution occurs at a total source-frame binary mass of $M_{\rm tot}\sim 40M_{\odot}$, with a reasonable range from $25M_{\odot}$ to $60M_{\odot}$, and a long tail that extends to higher masses.  This is consistent with the picture from \cite{Rodriguez2018}, where all GCs formed at $z\sim3.5$, and in-cluster mergers were limited to $\lesssim 60M_{\odot}$, since BHs more massive that $30M_{\odot}$ were ejected from the cluster in the early universe.  However, the young clusters in our cosmological model have not ejected all of their heavier BHs by the present-day, allowing these massive BHs to still participate in mergers in the local universe.  Of course, this effect is limited by the correlation between stellar metallicity and the maximum BH mass.  Clusters forming in the local universe have preferentially higher metallicities, and correspondingly lower maximum BH masses \cite[e.g.][]{Belczynski2010}.  In the bottom panel of Figure \ref{fig:masses}, we show the fraction of sources from each formation pathway as a function of the total mass.  As the mass of the BHs increases, the fractional contribution from ejected BBHs increases, since there exist very few low-metallicity young clusters that have retained their heavy BBHs up to the present day.  For the lower-mass BBHs, the contribution of ejected BBHs to the merger rate increases from roughly 50\% at $25M_{\odot}$ to nearly 100\% at $100M_{\odot}$.  The fraction of in-cluster mergers decreases over the same interval from 30\% to $\sim 10\%$, while the fraction of all mergers that occur from GW captures decreases from $\sim10\%$ to $\sim 5\%$.

This trend is more obvious in Figure \ref{fig:masses_1st}, where we focus only on the ``first-generation'' of BBH mergers.  In  this case, we have excluded the 15\% of BBH mergers which have components formed from the previous mergers of BBHs in the cluster.  The cutoff at $\sim 81M_{\odot}$ arises from the pair-production instability, where stellar core masses above $45M_{\odot}$ lose mass until they no longer undergo stellar pulsations, leaving behind BHs of at most $40.5M_{\odot}$ \cite{Belczynski2016a}.  Here, the fraction of ejected mergers increases smoothly from 50\% to 100\% from $25M_{\odot}$ to $100M_{\odot}$, while the in-cluster mergers decrease smoothly from 30\% to 20\% at $70M_{\odot}$, before dropping precipitously to $\sim 5\%$ at high masses.  Somewhat surprisingly, the fraction of GW capture events is relatively constant between $30M_{\odot}$ and $60M_{\odot}$, before decreasing to roughly 5\%.  For the heavier BBHs that merge in the local universe, the fraction that occur as isolated binaries versus GW captures are nearly equal.

\subsection{Mergers at $e\sim 1$}
\label{sec:pericenter}

The most striking feature in the eccentricity distribution of GW captures 
(Figure \ref{fig:ecc}) is the sharp peak at $e\sim1$.  This unique feature is 
consistent with previous scattering experiments 
\cite{Samsing2017,Samsing2018,Zevin2018}, and arises from the small fraction of 
BBH captures that form \emph{inside} the LIGO/Virgo detection band, with a peak 
GW frequency (equation \ref{eqn:wen}) greater than 10Hz.  We will now examine these systems in detail, which were set to $e\equiv0.99$ in the previous sections.

\begin{figure}[tb!]
\centering
\includegraphics[scale=0.85, trim=0in 0.in 0in 0.in, clip=true ]{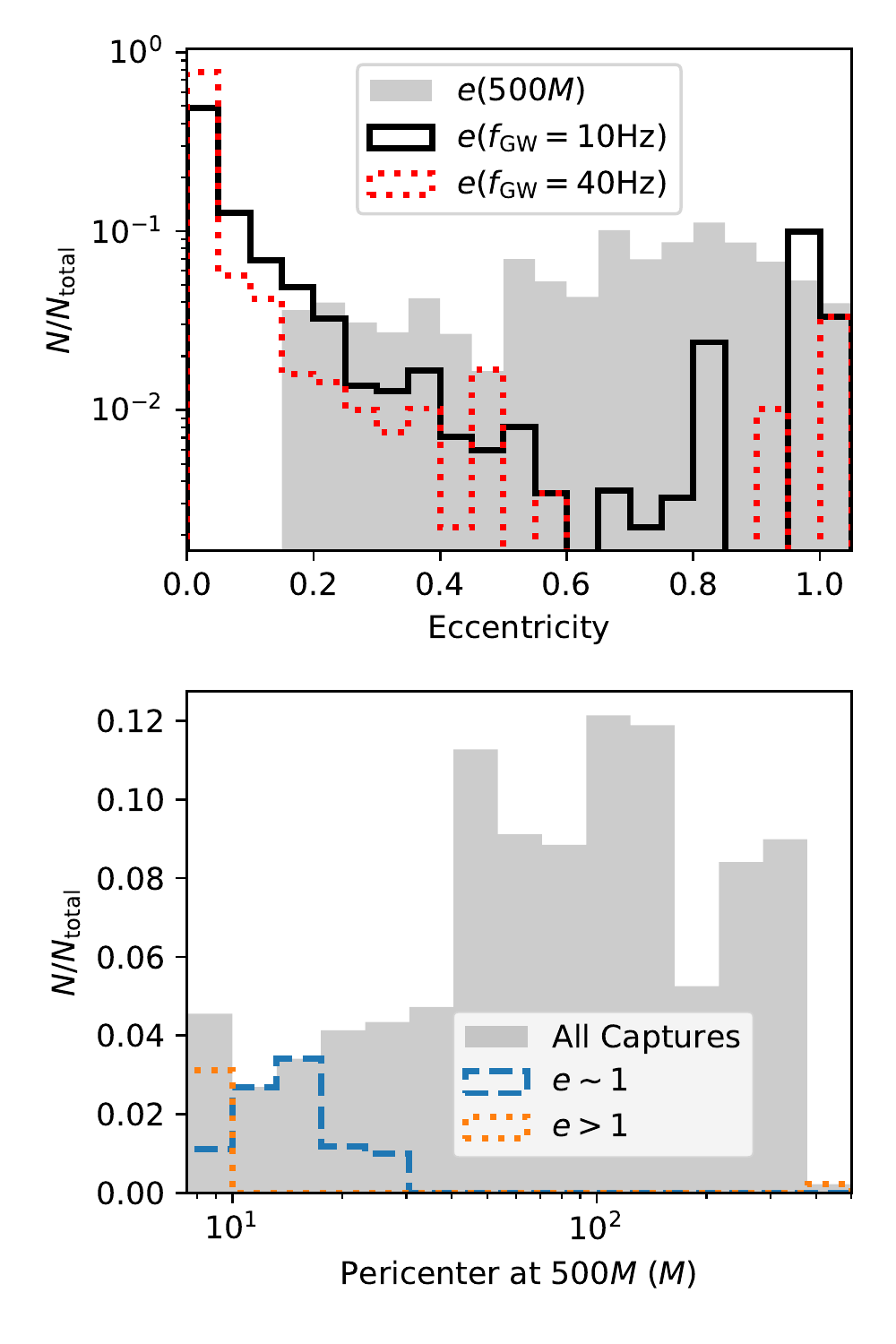}
\caption{The formation of the highly-eccentric GW capture events.  In the top plot, we show the measured eccentricities at a separation of $500M$ (solid gray), as well as the eccentricities once the binaries reach a peak GW frequency of 10Hz and 40Hz (in solid black and dotted red, respectively).  All of the binaries whose peak GW frequency is already in the LIGO band ($f > 10$Hz), which we record as having eccentricities of 0.99, evolve to lower, measurable values by 40Hz.  In the bottom panel, we show the pericenter separation for all the binaries at $500M$.  Those highly-eccentric captures with peak frequencies $\gtrsim10$Hz all form with very low pericenters distances, typically less than $30M$.  In both cases, the peak of systems with $e > 1$ are recorded as direct collisions.  In reality, these systems involved close pericenter passages of less than $10M$, which we cannot resolve using our pN scattering code.}
\label{fig:pericenter}
\end{figure}

In the top panel of Figure \ref{fig:pericenter}, we zoom in on the eccentricity 
distribution of the GW captures.  At an instantaneous separation of $500M$ (the largest separations reported by \texttt{Fewbody}), the distribution of 
eccentricities is biased towards very high values, with no GW captures passing 
through $500M$ with an eccentricity less that 0.1.  By the time these systems 
reach a peak GW frequency of $10{\rm Hz}$, the majority have been driven to 
$e<0.1$, with a tail extending to $e\sim0.6$.  However, there are three peaks at higher values
that bear mentioning.  First, at $e\sim0.8$, 
we see the beginning of the 
very-highly-eccentric mergers, where the systems form at such close pericenter 
distances that they cannot radiate away significant eccentricity before 
entering the LIGO/Virgo band.  The second peak, at $e=0.99$, represents BBHs which formed with peak GW frequencies greater than 10Hz.  Each of these sources form with very high eccentricities, which push the peak of GW emission to correspondingly high frequencies. As stated earlier, these sources do not formally have a peak frequency at 10Hz, as they formed with such high eccentricities that their peak frequencies are already greater than 10Hz.  However, all of these sources have peak frequencies of between 10Hz and 40Hz (when evaluated at $500M$).   If we instead ask what eccentricity each binary would have when it passes a peak GW frequency of 40Hz (the red dotted line in Figure \ref{fig:pericenter}), we find that all GW capture BBHs have well-defined eccentricities.

The third peak at $e \gtrsim 1$ are systems which were not bound at the time of merger.  Because we treat any BHs which pass within $10M$ of one another as merged, and because the pN approximation becomes unreliable at these separations, we cannot reliably track GW captures that form with 
pericenter separations of less than $10M$.  However, scattering experiments using 
the Monte Carlo models presented here as initial conditions \cite{Zevin2018}
have found almost no direct collisions occur between BHs when the Schwarzschild 
radius is used as the collision criterion.  These systems, which form with 
pericenter separations and velocities where the pN approximation breaks down, may 
retain significant eccentricities up to the point of merger.  However, since the 
energy emitted in GWs decreases as the eccentricities approach very large values \cite{Sperhake2008}, these systems are unlikely to produce observable GWs.

\section{Merger Rates}
\label{sec:rates}

In \cite{Rodriguez2018b}, we found that the total merger rate of BBHs from GCs was around 14 $\rm{Gpc}^{-3}\rm{yr}^{-1}$ at $z<0.1$.  This was calculated {by combining the models presented here} with the cosmological model for GC formation from \cite{El-Badry2018}, and is consistent with other recent estimates in the literature \cite[e.g.,][]{Fragione2018,Hong2018,Choksi2018a}.  Because of the distinct delay times between the in-cluster and ejected BBH mergers, we used separate phenomenological fits to the two populations, and found that the in-cluster mergers peaked earlier in redshift than BBHs that were ejected prior to merger.  For that work, the GW captures were classified as in-cluster mergers.  Of course, if the in-cluster mergers and GW captures follow different delay time distributions, it would be necessarily to separately fit all three populations.

\begin{figure}[tbh]
\centering
\includegraphics[scale=0.9, trim=0in 0in 0in 0.in, clip=true ]{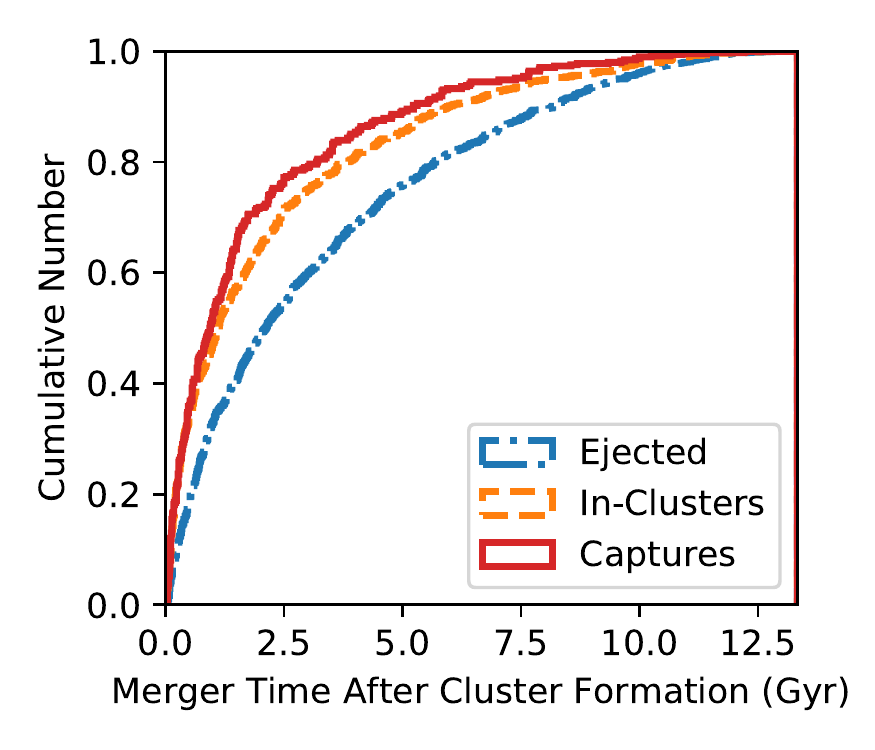}
\caption{The cumulative distribution of merger times for the ejected mergers, in-cluster mergers, and GW captures.  As expected, the ejected BBHs typically merge later due to the large delay times between ejection and merger that many systems can experience.  The in-cluster mergers and GW captures typically merge earlier, with nearly identical distributions.  Note that these distributions are for the merger times after GC formation, and have not been convolved with the cluster formation rate from Figure \ref{fig:formation}.}
\label{fig:delay}
\end{figure}

In Figure \ref{fig:delay}, we show the cumulative distribution of merger times for the ejected, in-cluster, and GW capture populations.  As was pointed out in \cite{Rodriguez2018b}, the ejected and in-cluster BBHs merge at preferentially different times; this is largely due to the additional delay time incurred by the ejected BBHs after their last encounter in the cluster, but before their eventual merger, which can be anywhere from many Myrs to many Gyrs later, depending on the escape speed of the cluster \cite[e.g.,][]{Rodriguez2016a}.

However, the distribution of merger times for the in-cluster and GW capture systems are nearly identical.  This is hardly surprising, given that both follow the evolution of the retained BH subsystem in the cluster, and depend strongly on the instantaneous encounter rate between BBHs and other stars/BHs.  Because of the nearly identical distributions, we calculate the rate of GW captures by multiplying the in-cluster merger rates from \cite{Rodriguez2018b} by the relative fraction of GW captures to all in-cluster mergers.

\begin{figure}[tb]
\centering
\includegraphics[scale=0.85, trim=0in 0in 0in 0.in, clip=false ]{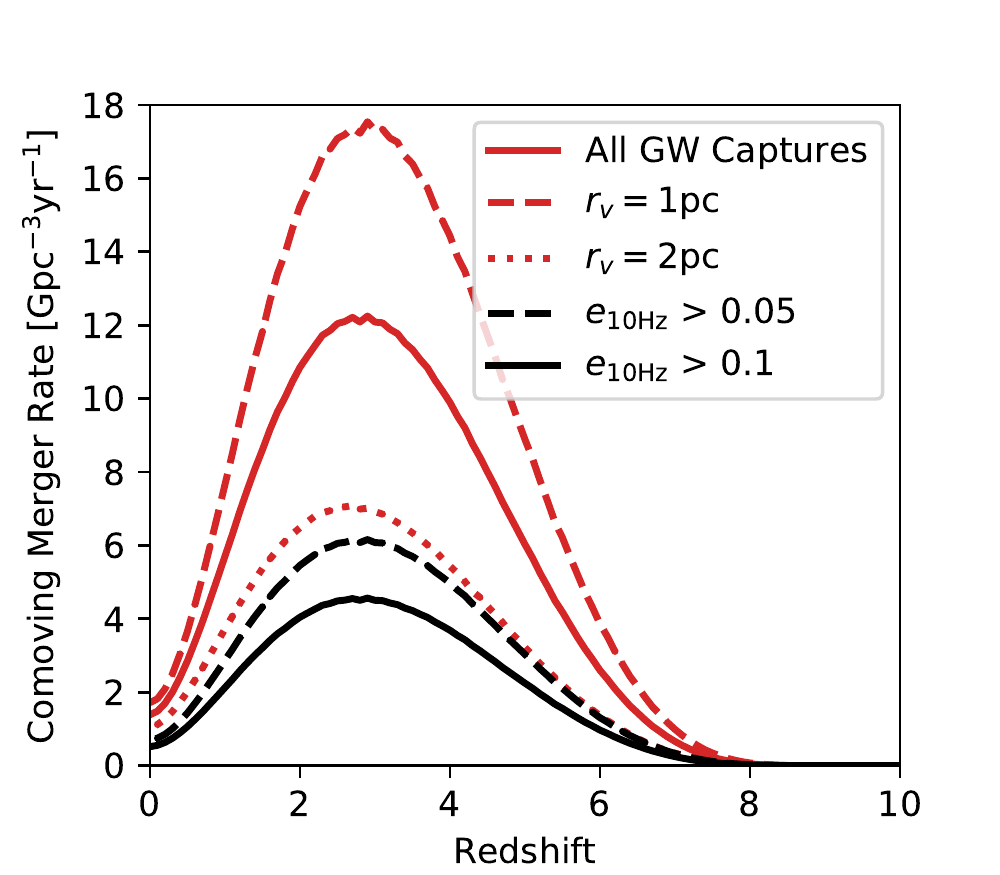}
\caption{The merger rate of GW captures across cosmic time.  Here, we use the merger rate from the in-cluster mergers of \cite{Rodriguez2018b}, normalized to the fraction of mergers that occur as highly-eccentric captures.  We show the merger rate of all GW captures in red, and show how the merger rate depends on the initial virial radii of the clusters.  In black, we show the fraction of these events which merge with eccentricities greater than 0.05 and 0.1 in dashed and solid black, respectively. }
\label{fig:rates}
\end{figure}

In Figure \ref{fig:rates}, we show the comoving merger rates for the GW captures as a function of redshift.  The solid red line shows the fraction of our standard model from \cite{Rodriguez2018b} that occur as GW captures, while the dashed and dotted lines show the merger rate if it were assumed that all clusters are born with initial virial radii of 1 or 2pc.  The total rate of GW captures varies from roughly 1 to 2 $\rm{Gpc}^{-3}\rm{yr}^{-1}$ at $z < 0.1$, with a peak anywhere from about 7 to 18 $\rm{Gpc}^{-3}\rm{yr}^{-1}$ at redshift 2.9 (2.7) for clusters with $r_v=1\rm{pc}$ ($r_v=2\rm{pc}$).  This decreases to 0.7 (0.5) $\rm{Gpc}^{-3}\rm{yr}^{-1}$ at $z<0.1$ when we restrict ourselves to the fraction of GW captures with measurable eccentricities of $e > 0.05$ ($e > 0.1$).  Note that these rates assumed the ``standard'' merger rate estimate of \cite{Rodriguez2018b}, which assumed a $1/M^2$ CIMF for clusters (as we have done here).  However, this rate is sensitive to the contribution from high mass clusters, and can decrease by a factor of 3 as the CIMF maximum mass is decreased from $10^7M_{\odot}$ to $10^6M_{\odot}$, which would further decrease the rates presented here.  At the same time, the fraction of GW captures does depend strongly on the contribution from binary-binary BBH encounters, which become more important at lower cluster concentrations and masses \cite{Zevin2018}.

Given that the LIGO/Virgo merger rates for BBH mergers in the local universe are anywhere from $32^{+33}_{-22}~\rm{Gpc}^{-3}\rm{yr}^{-1}$ (if a log-uniform BH mass function is assumed) to $103^{+110}_{-63}~\rm{Gpc}^{-3}\rm{yr}^{-1}$ (if a BH mass function following a $m^{-2.35}$ power law is assumed), the rate of highly-eccentric mergers is obviously not a dominant component of the total BBH merger rate.  Taking the upper and lower 90\% credible regions from the LIGO/Virgo rate as a bound, measurably eccentric GW captures may contribute anywhere from 0.25\% to 5\% of the total BBH merger rate.   While this represents only a small fraction of the total merger rate, the distinct eccentricities demonstrated here would provide a key discriminant between the many formation channels for BBH mergers.

\section{Conclusions}

In this paper, we have explored the formation and eccentricities of merging BBHs formed dynamically in GCs, with a particular focus on binaries that form during resonant encounters between BHs and BBHs, and taking into account the gravitational radiation reaction.  For the first time, we are able the study these systems in a fully cosmological context \cite{El-Badry2018} with a realistic population of GC models, whereas previous studies have been limited to isolated scattering experiments \cite[e.g.,][]{Gultekin2006,Samsing2018,Samsing2014,Samsing2017} or isolated models that were not representative of the observed population of clusters \cite[e.g.,][]{Rodriguez2018}.  We find that, when considering clusters with realistic initial conditions, GW captures contribute $12\%$ ($10\%$) of all BBH mergers from GCs at all redshifts (for redshifts $z<1$).  These mergers tend to have lower total masses than the BBHs that merge after ejection from the cluster, since the most massive BHs in the cluster were ejected many Gyr before the present day.  However, this trend is less pronounced when considering a realistic formation history of GCs across cosmic time, and becomes even weaker when multiple generations of BHs are allowed to form.

Combining the results of these simulations with the cosmological merger rate estimate from \cite{Rodriguez2018b}, we find that GW captures from GCs occur at a rate of 1-2 $\rm{Gpc}^{-3}\rm{yr}^{-1}$, in the local universe, increasing to a rate of 6 to 18 $\rm{Gpc}^{-3}\rm{yr}^{-1}$ at redshift 2.7 to 2.9, depending on what assumptions are made about the initial virial radius of the cluster.  When restricting ourselves to mergers that enter the LIGO/Virgo band with measurable eccentricities, we find local merger rates of 0.7 and 0.5 $\rm{Gpc}^{-3}\rm{yr}^{-1}$ for binaries that have eccentricities greater than 0.05 and 0.1, respectively.  While this represents a small fraction of the total LIGO/Virgo merger rate (thought to be between 10 and 213 $\rm{Gpc}^{-3}\rm{yr}^{-1}$ at 90\% confidence), the anticipated detection rate of one BBH merger per week for the upcoming LIGO/Virgo observing run suggests that the detection of a BBH with measurable eccentricity may be imminent.  Furthermore, proposed third-generation GW observatories, such as Cosmic Explorer \cite{Abbott2017f} and the Einstein Telescope \cite{Hild2011}, can potentially measure BBH mergers beyond redshift 10. Coupled with an anticipated increase in the sensitivity to lower-frequency GWs, we see that the merger rates presented in Figure \ref{fig:rates} may be directly measurable across cosmic time.  However, this will depend on how well the proposed GW detectors can measure orbital eccentricity: while some studies \cite{Lower2018} have suggested measurement accuracies as low as $e\sim 10^{-4}$, it has been noted that the parameter-estimation degeneracies between the BH spins and orbital eccentricities may complicate efforts to measure eccentricities below 0.1 \cite{Huerta2018}.  

While eccentricity is a clear indicator of dynamical processes in the formation of BBH mergers, there is significant work to be done to ascertain the eccentricity distributions associated with different dynamical channels.  There are several dynamical formation channels, including mergers from isolated field triples \cite{Antonini2017}, GW captures around super-massive BHs \cite{OLeary2009}, triples and captures formed in young open clusters \cite{Banerjee2017,Banerjee2018,Banerjee2018a}, and triples around SMBHs \cite{Antonini2014} that can produce BBH mergers with eccentricities detectable by LIGO/Virgo.  Further work will be needed to compare the global properties of these populations (such as the masses, spins, redshifts, and eccentricity distributions) to distinguish between these various dynamical scenarios.  Both the work presented here and recent work on GW captures in galactic centers \cite{Gondan2017a} have begun to probe the connection between the BH masses and eccentricities for some of these formation channels. 

We thank Emanuele Berti, Carl-Johan Haster, Scott Hughes, Cliff Will, and Nico Yunes for useful discussions.  CR
is supported by a Pappalardo Postdoctoral Fellowship at MIT.  This work was
supported by NASA Grant NNX14AP92G and NSF Grant AST-1716762 at Northwestern
University.  PAS acknowledges support from the Ram{\'o}n y Cajal Programme of
the Ministry of Economy, Industry and Competitiveness of Spain and the
COST Action GWverse CA16104.  CR and MZ thank the Niels Bohr
Institute for its hospitality while part of this work was completed, and
the Kavli Foundation and the DNRF for supporting the 2017 Kavli
Summer Program.  CR and FR also acknowledge support from NSF Grant PHY-1607611 to the Aspen
Center for Physics, where this work was started.

%\bibliography{mendeley_v2}
%\bibliographystyle{apsrev-carl}
\end{document}